
\magnification = 1200
\nopagenumbers
\overfullrule=0pt
\font\mmm=cmr9

\font\rmc=cmr9
\font\rmb=cmr9 scaled \magstep 1
\font\rma=cmr9 scaled \magstep 2
 3
\def\ueber#1#2{{\setbox0=\hbox{$#1$}%
  \setbox1=\hbox to\wd0{\hss$\scriptscriptstyle #2$\hss}%
  \offinterlineskip
  \vbox{\box1\kern0.4mm\box0}}{}}

\def\hn{$\;{\rm h}^{-1}$}
\def\R{\rm I\kern-.18em R}
\def\ref{\par\noindent\hangindent\parindent\hangafter1}
\def\etal{{\it et al. }}

\centerline{\rma HIGHER-ORDER LAGRANGIAN PERTURBATION THEORY}

\bigskip
\centerline{\rmb by}
\bigskip

\centerline{\rmb T. BUCHERT}

\bigskip

\centerline{\sl Max-Planck-Institut f{\"u}r Astrophysik}
\centerline{\sl 85740 Garching, Munich, Germany}

\bigskip

\centerline{ABSTRACT}
{\mmm
\baselineskip=10pt
{\narrower\medskip\noindent
Fundamental assumptions which form the basis of models for large-scale
structure in the Universe are sketched in light of a Lagrangian
description of inhomogeneities. This description is introduced
for Newtonian self-gravitating flows. On its basis a Lagrangian
perturbation approach is discussed and compared with
the standard Eulerian theory of gravitational instability.
The performance of Lagrangian perturbation solutions up to the
third order is demonstrated in comparison with
numerical N-body simulations. First results of this comparison are
presented for large scales (PM-code) and for small scales (tree-code).

\bigskip}}

\baselineskip=12.5pt
\rmc
\noindent
{\rma 1. Introduction}
\medskip\noindent
In conformity with the current status of observational insight into
the distribution of matter in the Universe, a model for the evolution
of matter inhomogeneities initiated by small-amplitude fluctuations
at the time of decoupling of matter and radiation has to cover scales
in the range
$\lbrack \lambda_{\max}, \lambda_{\min}\rbrack$, where $\lambda_{\max}$
is assumed to be considerably less than the horizon scale (today
about $6000$\hn Mpc), but larger than the `fair sample scale'
of a statistically representative volume of the Universe
(i.e., more than about $300$\hn Mpc, see Buchert \& Mart\'\i nez 1993),
and $\lambda_{\min}$ is assumed to represent the limit scale
down to which gravity rules the dynamics of matter (about $3$\hn Mpc).
Therefore, the
model cosmogony discussed below will be solely built on the
gravitational interaction;
the application range to the relevant spatial and temporal scales
will allow us to reduce the description to Newtonian
continuum mechanics of ``dust'' (i.e. pressureless) matter.
\medskip
The corresponding system of differential equations is the Euler-Poisson
system. We shall consider the Lagrangian representation of this
system, i.e., we represent
{\it comoving} Eulerian coordinates
$\vec q:=
\vec x / a(t)$ in terms of a vector field of Lagrangian
coordinates $\vec X$ which are constant along the flow-lines:
$$
\vec q = \vec F (\vec X,t) \;\;;\;\;\vec F (\vec X,t_i)=:\vec X\;\;,
\eqno(1)
$$
where $\vec x$ are coordinates of a non-rotating Eulerian frame,
$a(t)$ is the scale factor of a Friedman-Lema\^\i tre universe as a
homogeneous solution of the Euler-Poisson-system, here
specialized to an Einstein-de Sitter background ($a(t)=(t/t_i)^{2/3}$),
and $t_i$
denotes the time when initial data are given, e.g., after the time of
recombination (at the redshift $z_i = 1000$).
The Lagrangian evolution equations for the trajectory field $\vec F$
representing the Euler-Poisson system are to be found in (Buchert 1989,
1992, see also Ehlers \& Buchert 1993).
This system of equations is a closed system for the trajectories
alone from which all Eulerian fields like the velocity, the density,
the expansion, shear and vorticity of the flow, and also the
tensor of the gravitational tidal field which is a subject of
current discussion in the literature (Bertschinger \& Jain 1993),
can be computed.
\medskip
A Lagrangian cosmogony for the evolution of inhomogeneities is, in
its ideal form, an exact analytical solution of the Lagrangian
evolution equations for generic initial data which are given, e.g, in
terms of
Gaussian fluctuation fields for the density and the velocity. Since we
only know exact solutions for special
symmetry assumptions (globally plane or spherical symmetry (see, e.g.,
Peebles 1987),
or locally
plane symmetry (Buchert \& G\"otz 1987, Buchert 1989, Barrow \&
G\"otz 1989, Bildhauer \etal 1992, Silbergleit 1993)), the standard
technique to realize models is mostly provided by numerical N-body
simulations.
\medskip
More than two decades ago, Zel'dovich (1970, 1973) has realized the
limitations of the standard theory of gravitational instability which
makes use of the instability of the homogeneous solutions of the
Euler-Poisson system, and describes small-amplitude inhomogeneites
by solving the
system for small deviations of the density and velocity
fields in Eulerian space (see Peebles 1990, 1993 and ref. therein).
He designed an extrapolation of the linear Eulerian perturbation
solution into the non-linear regime by using Lagrangian coordinates.
His revival of an oldfashioned method proved to be very
successful as an analytical tool to model large-scale structures
as they were found in numerical simulations as well as observations, and
goes much beyond
the standard Eulerian solutions (see Coles \etal 1993). Although this
approximation has been widely used in the cosmological
community over-riding all what has been said from standard instability
scenarios, it was not generally
understood why this simple-minded extrapolation should be so successful.
\medskip
The investigation of the Lagrangian theory of gravitational
instability is refreshing to those who have begun to think that
a systematic self-explanatory derivation of Zel'dovich's
approximation is overdue: it can be viewed as a subclass of
irrotational Lagrangian first-order perturbation solutions (Buchert
1989, 1992).
We simply consider the Lagrangian perturbation
ansatz for the trajectory field $\vec f = \vec F / a(t)$ (Buchert 1992):
$$
\vec f = a(t) \vec X + \vec p \;\;;\;\;\vec p = \sum_m \varepsilon^m
{\vec p}^m \;\;, \eqno(2a)
$$
expand
the inhomogeneous deformation field $\vec p$ to a
given order $m$, and solve the Lagrangian evolution equations
up to this order. This approach yields solutions which contain
substantial non-linearities (see Section {\bf 2}),
and is therefore much better than the
standard perturbation approach in terms of deviations of the Eulerian
fields from the background cosmogony. One key-element for the
explanation of the success of this theory is that the density is
not a dynamical variable in the Lagrangian framework. As a consequence
the motion of
the continuum can be followed much further into the non-linear regime,
where high density excesses over the background density develop.
Indeed, the only restriction involved by this approach is the smallness
of the Lagrangian gradients of $\vec p$ compared with the
homogeneous background deformation; optimistically stated this
restriction reads:
$$
\vert \partial p_i / \partial X_k  \vert <
a(t)\;\;;\;\;a(t_i)=1\;\;.\eqno(2b)
$$
Lagrangian perturbation solutions derived in this framework are known up
to the third-order
(Buchert \& Ehlers 1993, Buchert 1993a,b).
The requirement of periodicity of $\vec p$ on the scale $\lambda_{\max}$
is, although used in all standard cosmogonies, not only a technical
advantage for the realization of the solutions; it is indispensible
to guarantee uniqueness of the solutions in Newton's theory
since otherwise there is considerable gauge freedom
(Ehlers \& Buchert 1993).
\medskip
Meanwhile, thanks to its success, the Lagrangian description has been
widely recognized as a powerful tool,
as is demonstrated by most recently exponentially
increasing
amount of work on Lagrangian (or mixed Lagrangian/Eulerian)
representations and solutions both
in Newton's theory (Moutarde \etal 1991, Matarrese \etal 1992,
Bouchet \etal 1992, Bernardeau 1992,
Lachi\`eze-Rey 1993a,b, Gramann 1993,
Bertschinger \& Jain 1993), and in
Einstein's theory which is intrinsically Lagrangian in the eigensystem
of the flow (Matarrese \etal 1993, Kasai 1993,
Croudace \etal 1993, see also Bertschinger \& Jain 1993).

\bigskip\bigskip
\smallskip
\noindent
{\rma 2. Comparison of Eulerian and Lagrangian perturbation theory}
\medskip\noindent
For the comparison of both perturbation approaches we look at
the equations governing the evolution of inhomogeneities
in the first-order Eulerian and Lagrangian approximations:
Consider the
contrast density $\Delta: = (\varrho - \varrho_H) / \varrho$,
$- \infty < \Delta < 1$,
which is more adapted to the non-linear situation than the conventional
definition $\delta=(\varrho - \varrho_H) / \varrho_H = \Delta /
(1 - \Delta)$, defined in Eulerian perturbation theory.
For this field we find the following fully non-linear
evolution equations
derived from the continuity equation and Poisson's
equation (Buchert 1989, 1992, see also Peebles 1987):
$$\eqalignno{
&\dot \Delta = (\Delta-1) {\bf I} \;\;, &(3a) \cr
&\ddot \Delta + 2 {\dot a \over a} \dot \Delta -
4 \pi G \varrho_H \Delta =
(\Delta-1) 2 {\bf II} \;\;,  &(3b) \cr}
$$
where ${\bf I}$ and ${\bf II}$
denote the first and second principal scalar invariants, respectively,
of the peculiar-velocity tensor gradient with respect to
comoving Eulerian coordinates $(\partial u_i / \partial
q_j)=(u_{i,j})$:
$$
{\bf I}:={1 \over a} \sum_i u_{i,i} \;\;\;,\;\;\;
{\bf II}:={1 \over 2 a^2} \left( (\sum_i u_{i,i})^2 -
\sum_{ij} \left(u_{i,j}u_{j,i}\right)\right)\;\;. \eqno (3c)
$$
For ${\bf II} = 0$ the equations (3) are
(except the Lagrangian time derivative and the redefinition of
the density contrast) the same equations as known in Eulerian linear
theory for
$\delta$ (Peebles 1980):
$$\eqalignno{
& {\partial \over \partial_t} |_q \delta^{\ell} = - {\bf I^{\ell}} \;\;,
& (3a)^{\ell} \cr
&{\partial^2 \over \partial_t^2} |_q \delta^{\ell}
+ 2 {\dot a \over a} {\partial \over \partial_t} |_q \delta^{\ell} -
4 \pi G \varrho_H \delta^{\ell} = 0\;\;. &(3b)^{\ell} \cr}
$$
These equations can be obtained by linearizing (3).
Contrary, the Lagrangian first-order solution solves the non-linear
equations (3) for ${\bf II} = 0$.
\medskip
I should stress that the similarity between the linear and the
restricted non-linear case is non-trivial:
An interesting excercise is
to compute the equation for the conventional density contrast $\delta$
from the non-linear equation
(3b) for ${\bf II}=0$ (solved by the Lagrangian first-order
solution) using the
definitions $\Delta:=\delta / (1 + \delta)$ and $\; \dot { }:=\partial_t
|_q + {\vec u \over a} \cdot \nabla_q$, and to compare with the
Eulerian linear equation (3b)$^{\ell}$:
\vfill\eject
$$
\ddot \delta \;+\; 2 {\dot a \over a} \; \dot \delta \;-\;
4 \pi G \rho_H \delta \;+
$$
$$
{\ddot \delta} \delta \;-\; 2 {\dot \delta}^2 \;+\;
2 {\dot a \over a} \; {\dot \delta} \delta \;-\;
4 \pi G \rho_H \delta^2 \;=\;0\;\;\;.\eqno(4)
$$
Hence, first-order Lagrangian perturbations
involve non-linearities
due to non-linearities in the dependent variable $\delta$, but also
due to products of $\vec u$ and $\delta$ involved by the Lagrangian
time-derivative.
\medskip
This exercise demonstrates the inherently
non-linear character of a Lagrangian perturbation approach.
Essentially, this property can be traced back to the implicit
solution of the Eulerian convection of the flow $(\vec u
\cdot \nabla_q)\vec
u$ by the Lagrangian time-derivative. A Lagrangian perturbation
approach solves for the deviations from trajectories while
moving with the perturbed flow, whereas in Eulerian space
particles soon experience large displacements from their
original positions implying large changes in the dependent variables.
\medskip
The second-order Lagrangian theory takes approximately
the term ${\bf II}$ in equation (3) into
account which covers essential effects of the tidal action on the fluid
and accelerates the collapse process significantly, whereas the
first-order model delays the collapse in comparison with numerical
simulations.
Also the local properties of the collapse on small scales are affected:
Second-order corrections contribute to the transformation of a
strongly anisotropic pancake collapse into a collapse to more prolate
objects (compare Buchert \& Ehlers 1993, Bertschinger \& Jain 1993)
which will be explained and illustrated below.

\bigskip\bigskip
\noindent
{\rma 3. Phenomenological and statistical performance of the solutions}
\medskip\noindent
As we already know from the analysis of Zel'dovich's approximation,
Lagrangian models perform well up to the epoch of first
shell-crossings in comparison with numerical N-body simulations, i.e.,
up to the intersection of trajectories
(see Coles \etal 1993). Therefore, all what will be said below
refers to properties around and after this epoch when differences
between first- and higher-order approximation schemes as well as
shortcomings of the Lagrangian perturbation approach in general become
evident.
\bigskip\noindent
{\rmb 3.1. Performance on small scales}
\medskip\noindent
The first striking improvement of higher-order schemes upon first-order
concerns the details of the collapse process of first
high-density objects. As was already mentioned the first-order scheme
delays the collapse significantly, first objects collapse much later
compared with spherically symmetric objects (see Blanchard \etal 1993
for a detailed discussion). Actually, an immediate consequence of
equation (3b) is that the gravitational collapse of a generic structure
occurs in any case faster than in the known spherically symmetric
solution (compare the discussion by Bertschinger \& Jain 1993).
It is therefore not surprising that the
second-order scheme predicts much faster collapse than the first-order
scheme, the spatial structure of high-density objects is more compact
due to tidal forces which add power to the motion along pancakes but
subtract power from the motion orthogonal to the pancakes. As a
consequence of enhanced gravitational potential within pancakes
($=$ three-stream systems of the flow), the inner trajectories don't
cross freely as in the first-order scheme, but recollapse into a second
shell-crossing. This feature is typical for self-gravitating flows
as is observed in N-body simulations; it is essential for the onset
of non-dissipative gravitational turbulence (a notion suggested by
Gurevich \& Zybin 1990 and ref. therein)
within pancakes (compare Buchert \& Ehlers 1993, Buchert 1993b, Buchert
\etal 1993b, and Figures 1,2,3).
\medskip
The third-order corrections redistribute the mass inside the
secondary mass-shells leading to a feature which appears as
{\it gravitational fragmentation} of mass-shells predicting compact
subclumps of matter inside the otherwise coherent structures (Fig.2).
If the original fluctuation has the size of superclusters (as in the
standard Hot-Dark-Matter model), then the subclumps have typical
extensions of large galactic halos. We can here only speculate that
higher-order corrections yield further shell-crossings inside this
hierarchy and develop further subclumps down to smaller and smaller
scales creating a cluster of `galactic halos' hierarchically in a
top-down fashion (i.e., from large to small scales).
Although the Lagrangian perturbation approach formally breaks
down just around the epoch of first shell-crossings, this study gives
interesting
clues to the general phenomenology of self-gravitating continua after
shell-crossing.
I emphasize that these features can only be seen in very high-resolution
realizations
possible in the analytical mapping. In view of these features, the
generally held view that
structure on small scales can only develop, if small-scale fluctuations
are present in the initial data is therefore wrong.
Analytical studies at high resolution provide reliable tools for
detecting such small-scale features, and it will be interesting to
analyze exact analytical solutions in order to fully understand
effects of gravitational fragmentation of pancakes.

\bigskip\noindent
{\rmb 3.2. Performance on large scales}
\medskip\noindent
While higher-order schemes add interesting details to the structures
on small scales, it is for cosmologists of prime interest to see how
well those approximations perform on large scales.
Coles \etal (1993) and Melott \etal (1993a)
conducted a series of tests by cross-correlating
density fields as modeled by N-body simulations with density fields
as predicted by the ``Zel'dovich approximation'' for various power
spectra. The methods used by Melott \etal where also used to
analyze the performance of the higher-order schemes by Buchert \etal
(1993a) and Melott \etal (1993d).
We found that the second-order scheme performs better than
the first-order scheme in the mildly non-linear regime (i.e. up to
the stage where the r.m.s. density contrast is of order unity) down
to the smallest resolved scale in the simulation. The
third-order scheme shows almost no improvement upon second-order
on those scales.
Our results also show that (at least for models which do not
involve much small-scale power in the initial conditions like the
standard Hot-Dark-Matter model, the Mixed-Dark-Matter models, or the
Isocurvature model),
the second-order Lagrangian perturbation solution shows excellent
agreement with the N-body simulations; these results carry
over to models
with more small-scale power (like the standard Cold-Dark-Matter model),
if we suitably truncate the initial data with high-frequency filters,
thus removing unwanted non-linearities. The use of Gaussian filters so
far
shows the best cross-correlations (see: Melott \etal 1993a,d,
Buchert \etal 1993a, Fig.4).

\vfill\eject
\noindent
{\rma 4. Problems with shell-crossing and future perspectives}
\medskip\noindent
All what has been said so far relies on an optimistic extrapolation
of solutions across caustics, i.e., regions of formally infinite
density in Eulerian space. In principle, the Lagrangian description
offers this possibility since the trajectory field as the only
dynamical variable remains regular, only the transformation to Eulerian
space is singular at points where the Jacobian of the transformation
degenerates. Formally this description is capable of modeling
multi-stream flow, i.e., internal structures of self-gravitating
pancakes, as can be appreciated in Figs.1,2. However, since the
gravitational field-strength in multi-stream regions has multiple
values in Eulerian space, a particle crossing the pancake experiences
a multiple of the gravitational force, whereas the Lagrangian flow
still follows the acceleration of individual particles. This fact
violates Einstein's principle of equivalence of inertial and
gravitational mass within multi-stream regions.
We conclude that the Lagrangian description formally allows to cross
caustics, but it implies neglection of the self-gravitating
interaction of streams.
This fact is responsible for the break-down of the Lagrangian models
discussed here and demands improvement.
\medskip
To circumvent this problem
various quasi-phenomenological approximations have been proposed,
among them the ``adhesion model'' (Gurbatov \etal 1989, Kofman \etal
1992) introducing a formal viscosity to mimic the gravitational
action of multi-stream flow, the ``frozen flow approximation''
(Matarrese \etal 1992) in which particles reach the first-generation
caustics asymptotically only, and the
``frozen potential approximation''
(Brainerd \etal 1993, Bagla \& Padmanabhan 1993) which
assumes approximate constancy of the gravitational
potential, while the density contrast can change by a huge factor.
However, all these models appear to be less accurate than the Lagrangian
approximation schemes concerning their dynamics, i.e.,
their cross-correlation with
N-body simulations in the fully developed non-linear regime
(Melott \etal 1993b,c), and their
higher moments of the density contrast -
skewness and kurtosis - in the weakly non-linear regime
(Bernardeau \etal 1993, Munshi \& Starobinsky 1993).
\medskip
In order to obtain a reliable model for self-gravitating multi-stream
systems one has to generalize
the framework in which structure formation is discussed. One possible
generalization is to study the Vlasov-Poisson-system of equations
for a distribution function of streams.

\bigskip
\noindent
{\bf Acknowledgements:}
\smallskip
\noindent
\noindent
I acknowledge financial support by DFG (Deutsche
Forschungsgemeinschaft).
\bigskip\bigskip
\noindent
{\rma References}
\bigskip
\ref
Bagla J.S., Padmanabhan T. (1993): {\it M.N.R.A.S.}, in press.
\ref
Barrow J.D., G\"otz G. (1989): {\it Class. Quantum Grav.} {\bf 6}, 1253.
\ref
Bernardeau F. (1992): {\it Ap.J.} {\bf 390}, L61.
\ref
Bernardeau F., Singh T.P., Banarjee B., Chitre S.M.
(1993): CITA preprint, SISSA bulletin board: astro-ph/9311055.
\ref
Bertschinger E., Jain B. (1993): {\it Ap.J.}, submitted.
\ref
Bildhauer S., Buchert T., Kasai M. (1992): {\it Astron. Astrophys.}
{\bf 263}, 23.
\ref
Blanchard A., Buchert T., Klaffl R. (1993): {\it Astron. Astrophys.}
{\bf 267}, 1.
\ref
Bouchet F.R., Juszkiewicz R., Colombi S., Pellat R. (1992):
{\it Ap.J.} {\bf 394}, L5.
\ref
Brainerd T.G., Scherrer R.J., Villumsen J.V. (1993): {\it Ap.J.}, in
press.
\ref
Buchert T. (1989): {\it Astron. Astrophys.} {\bf 223}, 9.
\ref
Buchert T. (1992): {\it M.N.R.A.S.} {\bf 254}, 729.
\ref
Buchert T. (1993a): {\it Astron. Astrophys.} {\bf 267}, L51.
\ref
Buchert T. (1993b): {\it M.N.R.A.S.}, in press.
\ref
Buchert T., Ehlers J. (1993): {\it M.N.R.A.S.} {\bf 264}, 375.
\ref
Buchert T., Mart\'\i nez (1993): {\it Ap.J.} {\bf 411}, 485.
\ref
Buchert T., Melott A.L., Wei\ss\ A.G. (1993a):
{\it Astron. Astrophys.}, in press.
\ref
Buchert T., Karakatsanis G., Klaffl R., Schiller P. (1993b):
{\it Astron. Astrophys.}, work in progress.
\ref
Coles P., Melott A.L., Shandarin S.F. (1993): {\it M.N.R.A.S.}
{\bf 260}, 765.
\ref
Croudace K.M., Parry J., Salopek D.S., Stewart J.M. (1993): {\it Ap.J.},
in press.
\ref
Ehlers J., Buchert T. (1993): in preparation.
\ref
Gramann M. (1993): {\it Ap.J.} {\bf 405}, L47.
\ref
Gurbatov S.N., Saichev A.I., Shandarin S.F. (1989): {\it M.N.R.A.S.}
{\bf 236}, 385.
\ref
Gurevich A.V., Zybin K.P. (1990): {\it Sov. Phys. JETP} {\bf 70}, 10.
\ref
Kasai M. (1993): {\it Phys. Rev.} {\bf D47}, 3214.
\ref
Kofman L.A., Pogosyan D., Shandarin S.F., Melott A.L. (1992):
{\it Ap.J.} {\bf 393}, 437.
\ref
Lachi\`eze-Rey M. (1993a): {\it Ap.J.} {\bf 407}, 1.
\ref
Lachi\`eze-Rey M. (1993b): {\it Ap.J.} {\bf 408}, 403.
\ref
Matarrese S., Lucchin F., Moscardini L., Saez D. (1992) {\it
M.N.R.A.S.} {\bf 259}, 437.
\ref
Matarrese S., Pantano O., Saez D. (1993): {\it Phys. Rev.} {\bf D47},
1311.
\ref
Melott A.L., Pellman T.F., Shandarin S.F. (1993a): {\it M.N.R.A.S.},
submitted.
\ref
Melott A.L., Lucchin F., Matarrese S., Moscardini L.
(1993b): {\it M.N.R.A.S.}, in press.
\ref
Melott A.L., Shandarin S.F., Weinberg D.H. (1993c):
{\it Ap.J.}, submitted.
\ref
Melott A.L., Buchert T., Wei\ss\ A.G. (1993d): in preparation.
\ref
Moutarde F., Alimi J.-M., Bouchet F.R., Pellat R., Ramani A.
(1991): {\it Ap.J.} {\bf 382}, 377.
\ref
Munshi D., Starobinsky A.A. (1993): preprint, SISSA bulletin board:
astro-ph/9311056.
\ref
Peebles P.J.E. (1980): {\it The Large Scale Structure of the Universe},
Princeton Univ. Press.
\ref
Peebles P.J.E. (1987): {\it Ap.J.} {\bf 317}, 576.
\ref
Peebles P.J.E. (1993): {\it Principles of Physical Cosmology},
Princeton Univ. Press.
\ref
Schiller P. (1992): {\it Ph.D.-Thesis}, LMU Munich (in German).
\ref
Silbergleit A.S. (1993): {\it Astron. Astrophys.}, submitted.
\ref
Zel'dovich Ya.B. (1970): {\it Astron. Astrophys.} {\bf 5}, 84.
\ref
Zel'dovich Ya.B. (1973): {\it Astrophysics} {\bf 6}, 164.

\bigskip\bigskip

\noindent
{\rma Figure Captions}
\bigskip
\noindent
{\bf Fig.1:} A simple two-dimensional cluster model clearly shows
the differences between first- and second-order Lagrangian
perturbation schemes (Fig.1a,b)
in comparison with a tree-code simulation
(Barnes-Hut-Schiller code; from: Schiller 1992)
for the same initial data (Fig.1e). We appreciate a second compact
structure inside the first collapsing pancake in the second-order
theory in excellent agreement with the tree-code simulation.
In Fig.1f
a spacetime section of the numerical
simulation of trajectories hitting two clusters of the same model
is shown based on a tree-code simulation.
This demonstrates the occurence of a hierarchy of shell-crossings
within the pancake due to the self-gravitating action of the
multi-stream system. A successively increasing number of streams
(3,5,7,{it etc.})
is observed resulting in a nested hierarchical structure
of mass-shells developing in the course of time.
Figs.1c,d show spacetime sections corresponding to the same model
here for the analytical schemes (1c: first-order,
1d: second-order, from: Buchert \& Ehlers
1993). We infer that the first-order solution describes the kinematical
formation of caustics, while the second-order solution
describes secondary
shell-crossings inside the first-generation pancakes.
\medskip
\noindent
{\bf Fig.2:} Three-dimensional realization of the third-order Lagrangian
perturbation
solution for a symmetric cluster model similar to that of Fig.1 at
high-spatial resolution
($1000^3$ trajectories collected into a $128^3$ pixel-grid).
Shown is a thin slice ($1/1000$ of the box) for an evolution
stage after the second shell-crossing.
A secondary filamentary structure develops inside the first pancakes
($=$ three-stream systems), and a second mass-shell is created inside
the cluster. The third-order corrections result in internal clustering
within secondary mass-shells and ``gravitational fragmentation''
of mass-shells; from: Buchert \etal (1993b).

\medskip
\noindent
{\bf Fig.3:} The variance $\sigma^2$ of the density contrast (smoothed
with rectangular top hats) for the model presented in Fig.1
is depicted versus the spatial scale variation at the stage
corresponding to the collapse time of the third-order model.
We infer that the first-order approximation starts to fail at
approximately the non-linearity scale (i.e., at $\approx 8$\hn Mpc
according to the standard normalization), while the
second-order scheme performs well down to scales of $\approx 3$\hn Mpc.
The third-order scheme reproduces the scale-invariance property of
a generic collapse (with slopes in the range ($-1.7,-1.8$)
as found in numerical simulations) down to the
smallest scale resolved.

\medskip
\noindent
{\bf Fig.4:} A comparison of the Lagrangian perturbation solutions
with numerical simulations (PM code by A.L. Melott)
is shown for the case of a hierarchical
model with considerable power on small scales (power-index $n=-1$).
Depicted is a thin slice of the density field ($1/64$ of the box).
Comparison is made for the numerical realization (Fig.4b)
(of the initial condition shown in Fig.4a)
(normalization as in Fig.3; box-size is $128$\hn Mpc),
and for the following analytical models:
Fig.4c: Eulerian linear perturbation solution; Fig.4d: Lagrangian
linear perturbation solution for the full spectrum, Fig.4e:
Lagrangian linear perturbation solution for the same spectrum truncated
at a scale corresponding to $\approx 2$Mpc\hn using a Gaussian filter,
and Fig.4f: same as Fig.4e but for the second-order scheme,
see: Melott \etal
(1993d). While the Eulerian solution merely enhances the initial
conditions in a self-similar way, the Lagrangian solutions are capable
to describe non-linear merging processes of small-scale structures.
Including perturbations below the truncation scale
results in `washing out'
small-scale structures. Using high-frequency filters, this shortcoming
can be repared by erasing unwanted
non-linearities which are not covered by the perturbation solutions.

\vfill\eject
\bye